\documentclass[footinbib,twocolumn,showpacs,amsmath,amstex,amssymb,mathfonts,superscriptaddress,prl]{revtex4}
\usepackage{graphicx}
\usepackage{color}
\usepackage{bm}
\usepackage{graphicx}
\usepackage{color}
\usepackage{amsmath}
\usepackage{amssymb}
\usepackage{amsthm}
\usepackage{amsfonts}
\usepackage{multirow}
\usepackage{makecell}
\usepackage{float}
\usepackage{comment}
\usepackage{enumitem}
\usepackage{verbatim}
\usepackage{mathtools}
\usepackage{esint}
\usepackage{tipa}
\usepackage{braket}
\usepackage{stackengine}

\newcommand{\obullet}[1]{\ensurestackMath{{\stackon[1.2pt]{#1}{{\mkern2mu\small\scriptstyle{\blacktriangledown}}}}}}
\newcommand{\ocirc}[1]{\ensurestackMath{{\stackon[1.2pt]{#1}{\mkern2mu\scriptstyle{\triangledown}}}}}
\renewcommand{\textsubring}[1]{\ensurestackMath{\stackon[-12pt]{#1}{\mkern2mu\circ}}}
\renewcommand{\d}[1]{\ensurestackMath{{\stackon[-11.5pt]{#1}{\mkern2mu\bf{{\color{red}\bf{{\scriptscriptstyle{\otimes}}}}}}}}}
\newcommand{\blankchar}{\color{white}{1}}

\usepackage{bbm}

\begin{document}

\title{Fractionalisation and dynamics of anyons at $\nu=n+1/3$ in fractional quantum Hall effect and their experimental signatures}
\author{Ha Quang Trung$^{1}$ and Bo Yang$^*$} 
\affiliation{Division of Physics and Applied Physics, Nanyang Technological University, Singapore 637371.}
\affiliation{Institute of High Performance Computing, A*STAR, Singapore, 138632.}
\email{yang.bo@ntu.edu.sg}


\date{\today}
\begin{abstract}
We show the low-lying excitations at filling factor $\nu=n+1/3$ with realistic interactions can be understood as quantum fluids with ``Gaffnian quasiholes" as the proper elementary degrees of freedom. 
Each Laughlin quasihole can thus be understood as a bound state of two Gaffnian quasiholes, which in the lowest Landau level (LLL) behaves like ``partons" with ``asymptotic freedom" mediated by neutral excitations acting as ``gluons". Near the experimentally observed nematic FQH phase in higher LLs, quasiholes become weakly bound and can fractionalise with rich dynamical properties. By studying the effective interactions between quasiholes, we predict a finite temperature phase transition of the Laughlin quasiholes even when the Laughlin ground state remains incompressible, and derive relevant experimental conditions for its possible observations.
\end{abstract}

\maketitle 

Strongly interacting topological systems in low-dimensions open doors to many exotic physics, particularly from the topological and geometric properties of low-lying excitations \cite{moore1991nonabelions,vaezi2014fibonacci,zhu2015fractional}. The fractional quantum Hall effect (FQHE) \cite{tsui1982two} is one such example, where a wide variety of topological phases can be realised from interactions between electrons under a strong magnetic field perpendicular to a two-dimensional manifold \cite{wen1992classification,wen1992theory}. A number of innovative techniques have been developed to understand both the universal topological properties and their dynamical robustness \cite{Laughlin1983,halperin1982quantized,Haldane1983,jain1989composite,jain1989incompressible,jain2007composite,Bernevig2008a,Yang2019}. These are all fundamentally non-perturbative approaches, because the kinetic energy of the system is completely quenched by the magnetic field, leaving behind only the effective interaction in a single Landau level (LL) \cite{Prange}.

Given that the electrons themselves are no longer good degrees of freedom, the challenge is to find suitable ``elementary particles" not perturbatively connected to electrons. Approaches along this line include the hierarchical pictures of the FQHE \cite{Laughlin1981quantized,Laughlin1983,halperin1982quantized,Haldane1983}, and later the composite fermion (CF) theory successfully explaining many experimental observations, especially in the lowest LL (LLL) for Abelian FQH states \cite{jain1989composite,jain1989incompressible,jain2007composite}. Extension to higher LLs (e.g. with the parton theory) and non-Abelian FQH states are also possible \cite{balram2013role,Balram2018a,balram2018fractional,Balram2020}, though they are technically more involved presumably because CFs also start interacting strongly. Alternative approaches with the Jack polynomial formalism\cite{Bernevig2008a,bernevig2008generalized,Yang2014} and the local exclusion constraint (LEC) as a generalisation\cite{Yang2019,Yang2019a,yang2019effective} seek a more microscopic understanding of the FQHE. There many universal topological properties of the FQHE can be determined algebraically without involving specific local operators (e.g. Hamiltonians). Interestingly, after identifying the model wavefunctions from this algebraic approach, we can in many cases construct model Hamiltonians for which the model wavefunctions are exact zero energy states \cite{Haldane1983, PhysRevLett.125.176402,simon2007generalized}. The algebraic approach is particularly useful in understanding non-Abelian FQHE, and has fundamental connections to the conformal field theory \cite{read2009conformal,Hansson2017}

In this Letter, we propose to understand and analyse the dynamics of the FQH phases using the quasiholes of apparently unrelated FQH phases, justified by the microscopic and algebraic relations revealed by the LEC construction \cite{Yang2019} and easily verifiable with numerical calculations. We focus on the familiar Laughlin phase at filling factor $\nu=n+1/3$, and predict rich dynamical phenomena with realistic interactions. In addition, the experimentally observed nematic FQH phase \cite{xia2011evidence} can be identified as the quantum critical point (QCP) separating the conventional Laughlin phase and a Haffnian-like phase. Near the QCP, Laughlin quasiholes can fractionalise into ``Gaffnian" \cite{simon2007construction} quasiholes carrying $e/6$ charge each. This is interesting given that the Gaffnian state is the subject of numerous studies due to its gapless nature \cite{simon2007construction,weerasinghe2014thin,kang2017neutral, jolicoeur2014absence,yang2021gaffnian}. Detailed analysis of the interactions between quasiholes predicts a finite-temperature quasihole phase transition with several experimental signatures.

{\it Low-lying excitations in the Laughlin phase --}
In the LLL we can understand the physics of the Laughlin phase from its model Hamiltonian (the $\hat V_1^{\text{2bdy}}$ Haldane pseudopotential). However, it is less clear what can happen when the Hall plateau is observed with realistic interactions far away from $\hat V_1^{\text{2bdy}}$. We first establish here for a wide range of interactions, the ground state and the low-lying excitations at $\nu=n+1/3$ live (almost) entirely within the Gaffnian quasihole (GQ) subspace. This subspace is defined algebraically with LEC condition $\{2,1,2\}\lor\{5,2,5\}$ \cite{Yang2019,Yang2019a}. It coincides with the null space (the GQs) of the Gaffnian model Hamiltonian $\hat H_g$, so we denote the subspace as $\mathcal H_G$. 

We demonstrate this by calculating the ground state of the 1LL Coulomb interaction $\hat V_{\text{1LL}}$, and show that its cumulative overlap with $\mathcal H_G$ is close to unity (see Table \ref{Laughlin overlap}). One should note in the thermodynamic limit, the GQ subspace is of measure zero in the full Hilbert space. Even for finite systems, it is quite nontrivial to have such high overlap decreasing slowly with system sizes within a subspace containing a small fraction of states. In contrast, the model wavefunctions for the Laughlin ground state and quasihole states have rather poor overlap to the true eigenstates of the interaction\cite{balram2013role,balram2020fractional,Balram2020}. 

\begin{table}
{\renewcommand{\arraystretch}{1.5}
\begin{tabular*}{\linewidth}{@{\extracolsep{\fill}}l c c c c c}
\hline
($N_e$,$N_o$) & (9,25) & (9,26) &(10,28) & (10,29) & (11,31) \\
\hline
$\mathcal{O}_{L}$ &0.48 & 0.45 & 0.54 & 0.47 & 0.70\\
\hline
$\mathcal{O}_{G}$ &0.97 &0.97 &0.97 & 0.89 & 0.97\\
\hline
dim($\mathcal{H}_G$)/dim($\mathcal{H}$) &0.143 & 0.135  &  0.091 & 0.077 & 0.039\\
\hline
\end{tabular*}
}
\caption{The overlap of the 1LL ground state with the Laughlin state, $\mathcal{O}^{L}=|\braket{\psi_{1LL}|\psi_{Laughlin}}|$ and the total overlap with the GQ subspace ($\mathcal{O}_{G}(\ket{\psi}) = \sqrt{\sum_{\ket{\phi}\in\mathcal{H}_G}|\braket{\psi_k|\phi}|^2}$) for the Laughlin ground states and one-quasihole states. The last row shows the dimension of the Gaffnian subspace used for calculation compared to the dimension of the full Hilbert space in the corresponding $L$ sector. }
\label{Laughlin overlap}
\end{table}

We can also look at the spectrum of $\hat V_{1LL}$ within $\mathcal H_G$, and find the low lying energies to approximate the exact energies from the full Hilbert space very well, unlike the variational energies from Laughlin model wavefunctions\cite{seesup}. The numerical evidence strongly suggests in a wide range of realistic interactions where interesting physics are observed at $\nu=n+1/3$, the low-lying excitations are quantum fluids made of ``GQs". At this filling factor, each GQ carries a charge of $e/6$ with respect to the Laughlin ground state (i.e. a charge of $e/5$ with respect to the Gaffnian ground state) \cite{seesup}. Thus a Laughlin quasihole (LQ) can be viewed as a bound state of two GQs.

{\it Dynamics of Gaffnian Quasiholes --}
We know that $\mathcal H_G$ is spanned by microscopic wavefunctions in the form of fermionic Jack polynomials $J_{\lambda}^{\alpha}\left(z_1,z_2,\cdots,z_{N_e}\right)$ with $\alpha=-3/2$ and root configurations $\lambda$ satisfying no more than two electrons for every five consecutive orbitals \cite{Bernevig2008a,bernevig2008properties,bernevig2009clustering}. The Gaffnian ground state has the root configuration $\lambda=1100011000\cdots 110001100011$. Here each digit corresponds to an electron orbital on the Haldane sphere\cite{Haldane1983}, with the left most digit corresponding the north pole and the right most digit the south pole. The digit ``1" implies the orbital is occupied by an electron, while the digit ``0" means the orbital is empty. 

Many physical properties of the state can be read off from the root configuration\cite{seesup}. The quasiholes can be created by inserting fluxes (or ``0's") to the ground state root configuration. Adding one flux creates two quasiholes, with two examples as follows:
\begin{eqnarray}
&&\textsubring{\blankchar}\textsubring{0}11000\cdots1100011\label{root1},\hspace{11pt}\textsubring{\blankchar}10100\cdots10100101\textsubring{\blankchar}\label{root2}
\end{eqnarray}
Here the positions of the quasiholes are marked by empty circles\cite{footnote4}. The two quasiholes can either form a bound state or be separated, given by the root configuration on the left and right in Eq.(\ref{root2}), respectively.

We now turn our attention to the Laughlin ground state, which has the root configuration:
\begin{eqnarray}
&&\textsubring{\blankchar}\textsubring{1}00\textsubring{1}00\textsubring{1}00\textsubring{1}00\cdots\textsubring{1}00\textsubring{1}00\textsubring{1}00\textsubring{1}\textsubring{\blankchar}\label{root3}
\end{eqnarray}
The Laughlin ground state can be seen as a rotationally invariant quantum fluid of GQs on the sphere. It has an excess of $N_e/2+1$ orbitals (where $N_e$ is the electron number) as compared to the Gaffnian ground state, implying the number of GQs it contains is $N_{qh}^G=N_e+2$. Adding one flux into (\ref{root3}) at the north pole yields one LQ as follows\cite{footnote4}:
\begin{eqnarray}
\d{{\ocirc{{\d{\blankchar}}}}}\textsubring{0}\textsubring{1}00\textsubring{1}00\textsubring{1}00\textsubring{1}00\cdots\textsubring{1}00\textsubring{1}00\textsubring{1}00\textsubring{1}\textsubring{\color{white}1}\label{root4}
\end{eqnarray}
Here the LQ is denoted by the empty triangle. Recall that adding one flux is also equivalent to adding two GQs - the \emph{additional} GQs are denoted by red crossed circles. In this configuration, the two GQs are on top of each other, forming the bound state at the north pole that is the LQ. 

Splitting the GQ pair can be achieved by violating the admissibility rule of the Laughlin state, while still satisfying the admissibility rule of the Gaffnian state. One example is given as follows:
\begin{eqnarray}
&&\d{\blankchar}{\ocirc{{\textsubring{\blankchar}}}}\textsubring{0}1{\obullet{{0}}}1\textsubring{0}{\ocirc{{0}}}\textsubring{0}1{\obullet{{0}}}1\textsubring{0}{\ocirc{{0}}}\textsubring{0}\cdots1{\obullet{{0}}}1\textsubring{0}{\ocirc{{0}}}\textsubring{0}1{\obullet{{0}}}1\textsubring{0}\ocirc{{{\textsubring{\blankchar}}}}\d{\blankchar}\label{root7}
\end{eqnarray}
\begin{figure}
\begin{center}
\includegraphics[width=\linewidth]{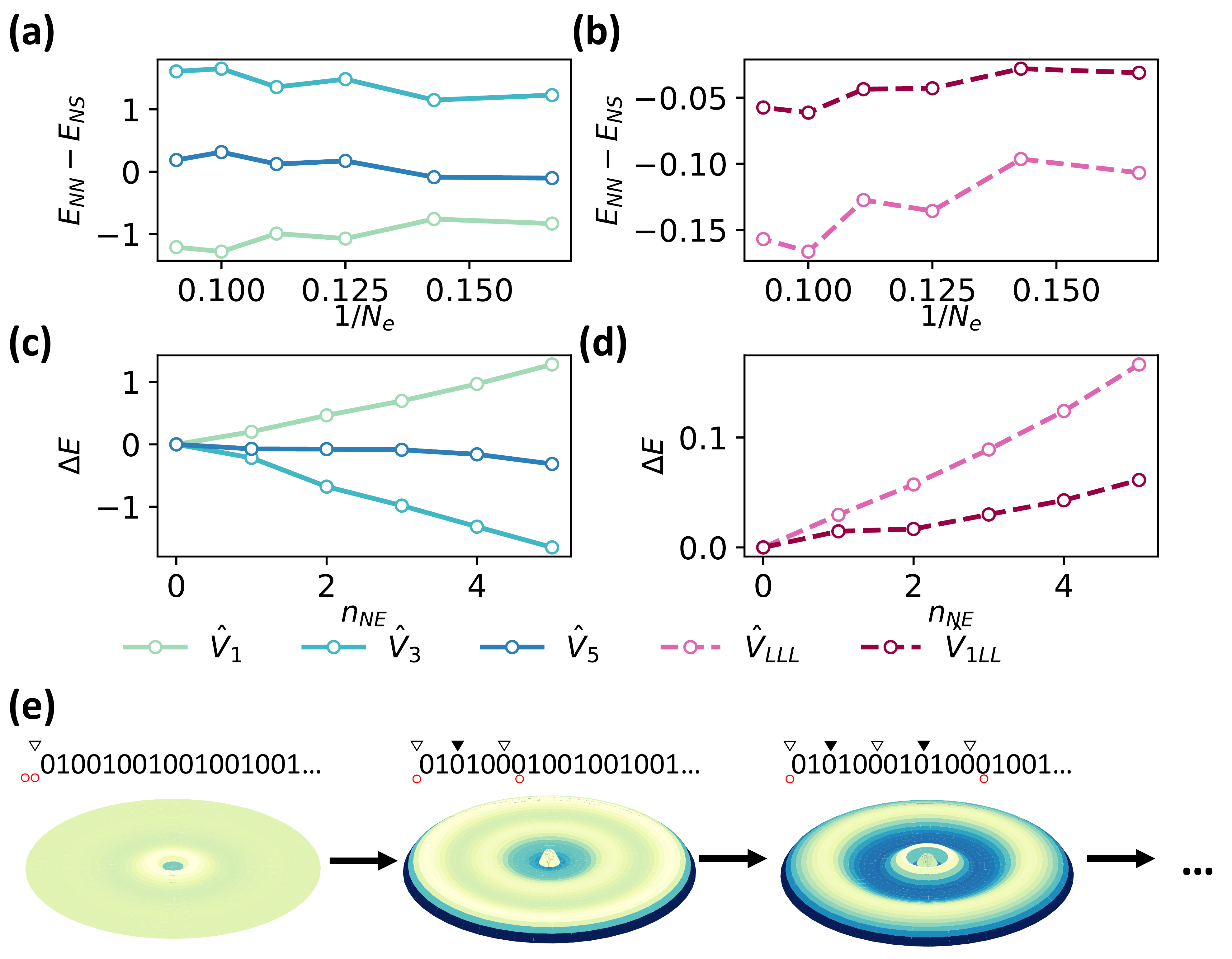}
\caption{(a-b) Variational energy difference between bound ($E_{NN}$) and unbound ($E_{NS}$) GQs, plotted against system size. (c-d) The energy cost to separate the two GQs plotted against the number of neutral excitations between them, for systems with 10 electrons and 28 orbitals. (e) Electron density of the quasihole wavefunctions. The the center of the disk corresponds to the north pole of the sphere. The leftmost Laughlin state has a bound quasihole at the center, shown by the dip in electron density. As the GQs are pulled apart, neutral excitations are formed, shown by the ripples around the center.}
\label{fig1}
\end{center}
\end{figure}
The solid triangles give the positions of Laughlin quasiparticles\cite{footnote4}. Given any microscopic Hamiltonian, the variational energies of the corresponding many-body states show if the two GQs are attractive or repulsive. We can clearly see that such interaction is mediated by neutral excitations, or LQ-quasiparticle pairs. Thus with $\hat V_1^{\text{2bdy}}$, the GQs are strongly attractive with the interaction energy \emph{increasing} with the quasihole separation, mimicking the ``asymptotic freedom" for quark system where the neutral excitations play the role of ``gluons". With $\hat V_3^{\text{2bdy}}$, however, GQs are repulsive (see Fig. \ref{fig1}). Thus with realistic interactions, the GQs can be either bound, weakly bound or unbound.

{\it Nematic FQH state as the quantum critical point --}
Let us first compare the Laughlin model state in Eq.(\ref{root3}) with the Haffnian model state with the following root configuration\cite{green2001strongly,hermanns2011irrational}:
\begin{eqnarray}
&&110\textsubring{0}\textsubring{0}0110\textsubring{0}\textsubring{0}0\cdots 110\textsubring{0}\textsubring{0}011\label{root8}
\end{eqnarray}
Both the many-body states of Eq.(\ref{root3}) and Eq.(\ref{root8}) are zero energy states of $\hat H_g$, and are thus linear combinations of the Gaffnian Jack polynomials. From the reduced density matrix near the north pole, we can see that the Laughlin state is made of bound GQs, but the Haffnian state is made of unbound GQs. From Fig.(\ref{fig1}) we thus expect the Haffnian state to have \emph{extensive} variational energy with respect to $\hat V_1^{\text{2bdy}}$.

We can now have a better understanding of the dynamics of the magnetoroton modes for the Laughlin phase at $\nu=1/3$, with the following root configurations \cite{yang2012model,Yang2014}:
\begin{eqnarray}
&&110\textsubring{0}\textsubring{0}\textsubring{0}\textsubring{1}00\textsubring{1}00\textsubring{1}00\cdots\qquad L=2\\
&&1100\textsubring{0}\textsubring{1}\textsubring{0}0\textsubring{0}\textsubring{1}00\textsubring{1}00\cdots\qquad L=3\label{dipole}\\
&&\qquad\qquad\vdots\nonumber
\end{eqnarray}
where $L$ is the total angular momentum quantum number on the sphere. These are no longer Jack polynomials, but we know immediately from LEC that the entire branch of the magnetoroton mode lives in $\mathcal H_G$ (i.e. zero energy states of $\hat H_g$). The quadrupole excitation at $L=2$ is also the zero energy state of the Haffnian Hamiltonian $\hat H_h$ \cite{Yang2020s}. With the $\hat V_1^{\text{2bdy}}$ interaction, the quadrupole excitation has higher energy because it consists of an unbound pair of GQs. In contrast, the dipole excitations in the limit of large $L$ define the incompressibility gap and consist of LQs as bound GQs. Thus the energy difference between quadrupole and dipole excitations results from the $\hat V_1^{\text{2bdy}}$ favouring bound states of GQs.

\begin{figure}
\begin{center}
\includegraphics[width=\linewidth]{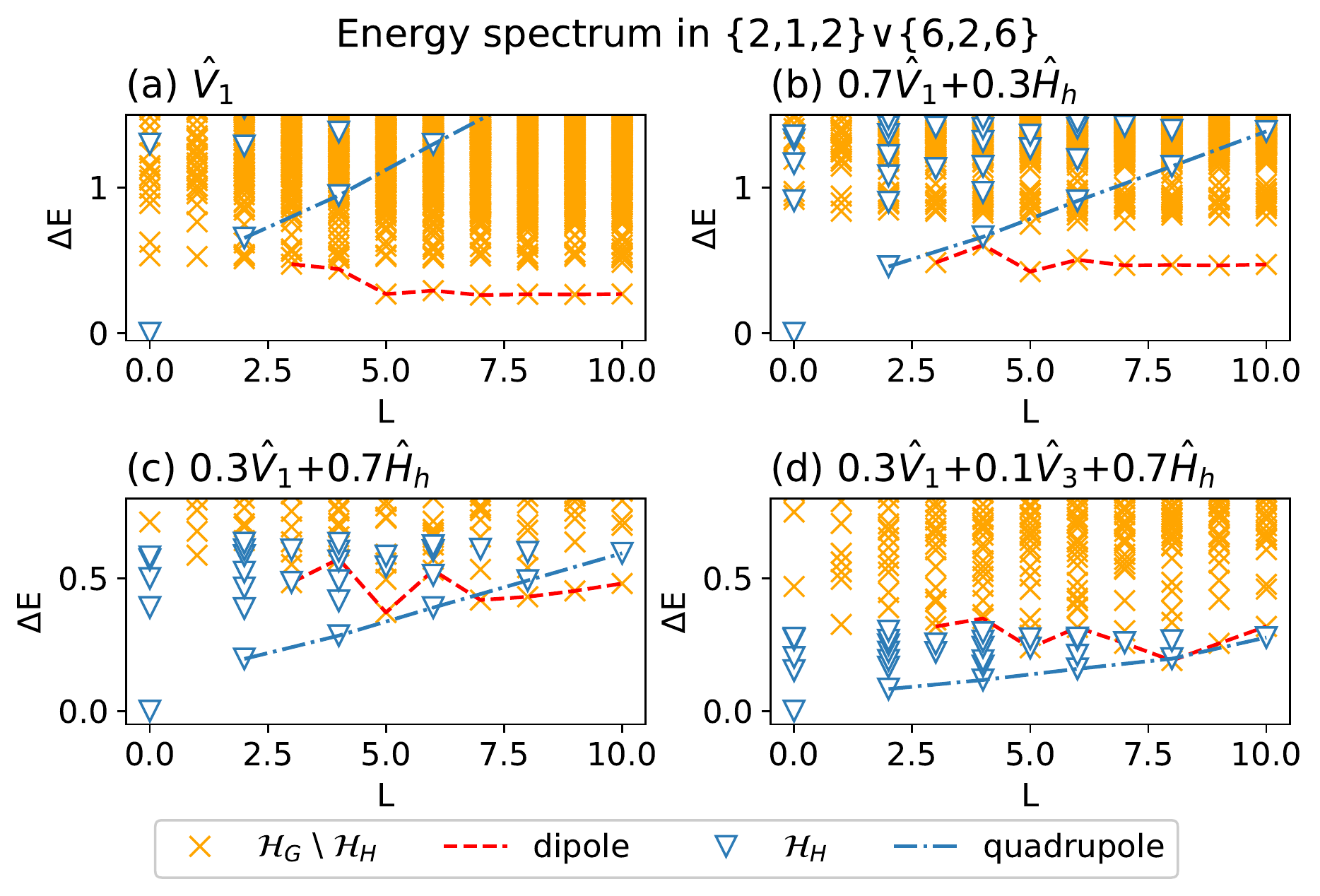}
\caption{Energy spectrum for system in the Haffnian subspace (blue triangles) and the complement of the Haffnian subspace in the Gaffnian subspace (orange crosses). The dipole and quadrupole excitation branches are highlighted in red and blue dotted lines, respectively. Results on a system with 10 electrons and 28 orbitals evaluated is shown here. The important features of this numerics are robust and consistent with the full ED spectrum\cite{seesup}.}
\label{fig2}
\end{center}
\end{figure}

The nematic FQH state, an experimentally observed phase where the quantum Hall plateau coexists with the anisotropic longitudinal transport at low temperature \cite{xia2011evidence, feldman2016observation}, is believed to result from the quadrupole excitations going soft at $\nu=2+1/3$\cite{you2014theory,regnault2017evidence}. Its underlying microscopic mechanism, however, is still not fully understood \cite{regnault2017evidence,Yang2020s}. Here, we show that the quadrupole excitations going soft results from Hamiltonians favouring unbound GQ pairs. The physics can be captured by the following toy model \emph{within} $\mathcal H_G$:
\begin{eqnarray}\label{model}
\hat H\left(\lambda_1,\lambda_2\right)=\hat H_{\text{h}}+\lambda_1\hat V_1^{\text{2bdy}}+\lambda_2\hat V_3^{\text{2bdy}}
\end{eqnarray}
Given the assumption that $\hat V_1^{\text{2bdy}}$ is incompressible at $\nu=1/3$, we know $\hat H\left(\lambda_1,0\right)$ is also incompressible for any positive $\lambda_1$. With $\lambda_2=0$ and small $\lambda_1$, the dipole excitations (and thus the charge excitations) are gapped by both $\hat H_{\text{h}}$ and $\hat V_1^{\text{2bdy}}$. The lowest energy excitation is given by the quadrupole excitation in the $L=2$ sector. A ``linear" dispersion in the even $L$ sector can be seen (Fig.(\ref{fig2})). They correspond to the multiple quadrupole excitation states with unbound GQs, with the following root configurations\cite{footnote1}:
\begin{eqnarray}
&&110000100100100100100100\cdots\qquad L=2\label{g1}\\
&&110000110000100100100100\cdots\qquad L=4\label{g2}\\
&&\qquad\qquad\vdots\nonumber
\end{eqnarray}
Let us denote the neutral gap of the system to be $\Delta_n$, and the charge gap to be $\Delta_c$ (both with respect to the lowest energy state in the $L=0$ sector). The nematic FQH is realised in the regime of $\Delta_n\ll \Delta_T\ll\Delta_c$ as given by the $\lambda_1\ll 1,\lambda_2=0$ model, where $\Delta_T$ is the energy scale of the temperature or disorder. We can also make the nematic FQH phase more robust by increasing $\lambda_2$. This is because $\hat V_3^{\text{2bdy}}$ punishes the Laughlin ground state and dipole excitations (they consist of bound GQs), while energetically favouring the quadrupole excitation (see Fig.(\ref{fig2}d)).

By increasing $\lambda_2$ from zero, we enter the regime where the quadrupole excitation becomes gapless. Here the dispersion of multiple quadrupoles becomes truly linear\cite{footnote2} in the long wavelength limit with an effective velocity $v_g$. At the QCP, $v_g=0$, implying the Laughlin state and the Haffnian state become degenerate. We thus expect that tuning of $\lambda_2$ allows us to access a Haffnian-like phase at $\nu=1/3$ with topological shift $S=-4$ (in contrast to $S=-2$ for the Laughlin phase)\cite{footnote3}.

While Eq.(\ref{model}) is artificial, it is actually more realistic than it appears. The mean-field interaction $\hat H^{\text{2bdy}}_{\text{h}}=\hat H_{\text{h}}+\hat H^*_{\text{h}}$, where $\hat H^*_{\text{h}}$ is the particle-hole conjugate, is a rather physical short range two-body interaction (consisting of $\hat V_1^{\text{2bdy}},\hat V_3^{\text{2bdy}},\hat V_5^{\text{2bdy}}$). In the thermodynamic limit, replacing $\hat H_{\text{h}}$ with $\hat H^{\text{2bdy}}_{\text{h}}$ in Eq.(\ref{model}) may retain the qualitative features of the original model\cite{sreejith2017surprising,kusmierz2019mean}. In higher LLs, three-body interactions also arise from LL mixing, which can play an important role to the physics near the QCP \cite{sodemann2013landau}. 

In Fig.(\ref{fig2}) we show the transition between the dipole excitations and the quadrupole excitation when increasing $\lambda_1,\lambda_2$. While the mixing of the two branches of excitations complicates the dynamics of the neutral excitations, the softening of the quadrupole excitation at $L=2$ sector is not affected, since there is no competing single-dipole excitation (which starts at $L=3$) in this sector. The condition of $\Delta_n\ll\Delta_c$ is also maintained for a large parameter range.

Unlike the dipole excitations, states containing quadrupole excitations (e.g. Eq.(\ref{g1}) and Eq.(\ref{g2})) have uniform electron density in the thermodynamic limit. They are thus conjectured to be the nematic Goldstone mode proposed in the effective field theories \cite{you2014theory,maciejko2013field}. Thus for the effective field theory to be relevant to the nematic FQH, the microscopic interaction has to gap out all states \emph{not} in $\mathcal H_G$. The effective theory also assumes $\Delta_n<0<\Delta_c$ with $v_g>0$ in the nematic phase. Microscopically, since the ground state energy, quadrupole and dipole energies are fundamentally determined by the dynamics of GQs, they cannot really be independently tuned. The likely scenario relevant to the experiments is for $0<\Delta_n<\Delta_T<\Delta_c$ with $v_g\sim\Delta_n l_B$. For $\Delta_n<0$ we are no longer in the Laughlin phase and may not have a charge gap with $v_g<0$ .

{\it Experimental signatures of quasihole fractionalisation--}
Unlike $\hat V_1^{\text{2bdy}}$ or $\hat V_{\text{LLL}}$, the interaction close to the nematic FQH no longer heavily punishes unbound GQs. This can also been seen from the root configuration of an unbound pair of GQs in Eq.(\ref{root7}). The interaction between them is mediated by quadrupole-like neutral excitations (satisfying LEC condition $\{2,1,2\}\lor\{6,2,6\}$), instead of the dipole-like ones (satisfying $\{2,1,2\}\lor\{5,2,5\}$). Indeed, Eq.(\ref{model}) can serve as the model Hamiltonian both for the quasihole fractionalisation and the nematic FQH transition\cite{footnote5}. Thus near the nematic FQH phase, $\Delta_n\ll\Delta_c$ implies an incompressible phase given by $\Delta_c$ and thermally excited quasiholes of charge $e/6$ (with excitation energies given by $\Delta_n$). Unlike individual quarks confined to very small length scales and are thus unobservable, at the Laughlin phase $e/6$ GQs are bound at the order of a few magnetic lengths. Near the QCP and at finite temperature the separation of GQs can be significantly larger. We thus expect $e/6$ charge to be observable in the bulk with single electron tunnelling experiments\cite{beenakker1993single} near the nematic FQH phase.

\begin{figure}
    \centering
    \includegraphics[width=\linewidth]{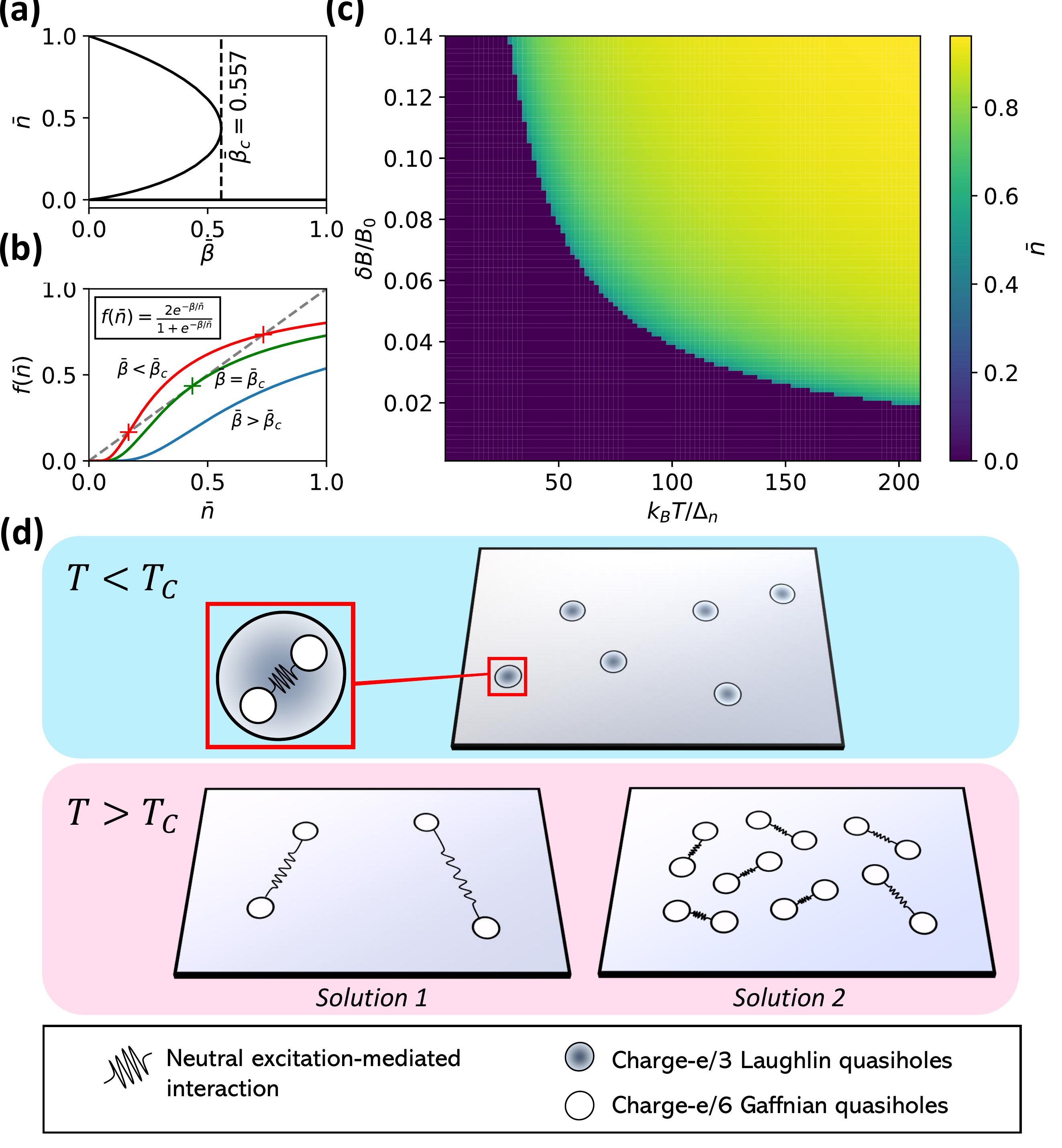}
    \caption{(a) $\bar{n}$ as a function of reduced temperature $\bar\beta$. (b) Solutions of Eq.(\ref{condition}) are marked with crosses. At $\bar\beta_c$ there is a unique solution. (c) $\bar n$ as a function of $\delta B$ and temperature. (d) when $T<T_C$, only LQs are present in the system; when $T>T_C$, there can be a lower GQ density solution (each GQ pair has a higher energy cost), and a higher unbound GQ density solution(with shorter average separation distance, hence lower energy cost for each GQ pair).} 
    \label{fig3}
\end{figure}

We also predict a quasihole phase transition at the incompressible Laughlin phase at some critical temperature, similar to the Berezinskii–Kosterlitz–Thouless (BKT) transition\cite{kosterlitz1973ordering,thouless1998topological}, evidenced by the softening of the quadrupole modes in Fig.(\ref{fig2}). Let $n,n_{\frac{e}{6}}$ be the density of additional magnetic flux to the ground state, and of GQs with charge $e/6$, respectively. The average distance between any two GQs is thus $\bar d\sim 1/\sqrt{n_{\frac{e}{6}}}$. From Eq.(\ref{root7}) and the linear dispersion in Fig.(\ref{fig3}), the average energy cost is proportional to the number of quadrupole excitations between two fractionalised GQs, with $\Delta E\simeq \bar{\Delta}_n\bar d^2,\bar{\Delta}_n=\Delta_n/\left(3\pi^2l_B^2\right)$. We can thus define dimensionless quantities $\bar n=n_{\frac{e}{6}}/n$ and $\bar\beta=\beta n^{-1}\bar{\Delta}_n, \beta=\left(k_BT\right)^{-1}$, satisfying the following\cite{seesup}:
\begin{eqnarray}\label{condition}
\bar n\left(1+e^{\bar{\beta}/\bar n}\right)=2
\end{eqnarray}
There is thus a critical temperature given by $\bar\beta_c=0.55693$. The $\bar n>0$ solutions, implying a finite density of $e/6$ quasiholes, only exist for $T>T_c$ with the following:
\begin{eqnarray}
\label{Tc}
T_c=T_n\frac{2}{3\pi\bar\beta_c}\left(\frac{\delta B}{B_0}\right)^{-1}\sim 0.381 T_n\left(\frac{\delta B}{B_0}\right)^{-1}
\end{eqnarray}
where $T_n=\Delta_n/k_B$ is the quadrupole gap temperature; $B_0$ is the magnetic field at the center of the $\nu=n+1/3$ plateau, and $\delta B$ is the deviation of the magnetic field from $B_0$ on the quasihole side. The solutions to Eq.(\ref{condition}) and the intuitive picture of this phase transition are illustrated in Fig.(\ref{fig3}).

In experiments, lower $T_c$ is preferred because the $e/6$ quasiholes are only observable if $k_BT_c$ is smaller than the charge gap. We see from Eq.(\ref{Tc}) that this can be done by lowering $\Delta_n$, and from Fig.(\ref{fig2}) that $\Delta_n$ can be lowered by adding long-range interaction ($V_3^{2bdy}$). Thus, we expect such a window to exist in higher LL and near the nematic FQH phase, where $\Delta_n\sim 0$ can be potentially realized in experiments \cite{xia2011evidence, feldman2016observation, you2014theory,regnault2017evidence}. Finally, we emphasize that the robustness of the Hall plateau at $\nu=n+1/3$ in experiments does not automatically imply well quantised quasiparticle charge of $e/3$ in shot noise or tunnelling experiments.

\begin{acknowledgments}
We are grateful to A. Balram and X. Lin for helpful discussions. This work is supported by the Singapore National Research Foundation (NRF) under NRF fellowship award NRF-NRFF12-2020-0005.
\end{acknowledgments}

\nocite{*}
\bibliographystyle{apsrev}
\bibliography{ref}

\clearpage

\renewcommand{\thefigure}{S\arabic{figure}}
\renewcommand{\theequation}{S\arabic{equation}}
\renewcommand{\thepage}{S\arabic{page}}
\setcounter{figure}{0}
\setcounter{page}{0}

\onecolumngrid
\begin{center}
{\large \textbf{Online supplementary material for ``Fractionalisation and dynamics of anyons at $\nu=n+1/3$ in fractional quantum Hall effect and their experimental signatures"}}

\mbox{%
  \parbox{0.78\textwidth}{
In the first part of this supplementary, we elaborate on how the Laughlin state can be viewed as a fluid of Gaffnian quasiholes (GQs), and illustrate the splitting of the Laughlin quasihole (LQ) by contrasting the electron density obtained from the Jack polynomial corresponding to the split GQ pair with that of the Laughlin wavefunction. Next, we present the spectrum obtained from full ED of systems near the QCP. The subsequent sections present detailed calculation on the separation energy of the GQ pair, and the temperature dependence of the GQ density in a system.
  }%
}
\end{center}

\twocolumngrid
\section{S1. Root configuration - binary representation and model wavefunctions}
In this section we describe in details the connection between the binary root representation and its corresponding model wavefunction. All our discussions are applicable on the Haldane sphere, as well as the disk or cylinder geometry. The description of a wavefunction based on its root configuration forms the basis for our physical arguments in the main text. 

Many FQH states can be expressed as Jack polynomials\cite{Bernevig2008a,bernevig2008generalized}, which are well-defined multivariate polynomial parametrized by a \emph{partition} called root configuration\cite{macdonald1998symmetric}. There are also FQH model wavefunctions, such as the Haffnian\cite{green2001strongly}, that do not belong to the family of Jack polynomials yet are still captured by a root configuration. The LEC formalism\cite{Yang2019} captures both the Jack polynomial states and many of these other states - in that sense, the LEC is a generalization of the Jack formalism. This section reviews and defines the general concepts of ``partitions", ``root configuration" and ``monomials".

\begin{figure*}
\begin{center}
\includegraphics[width=\linewidth]{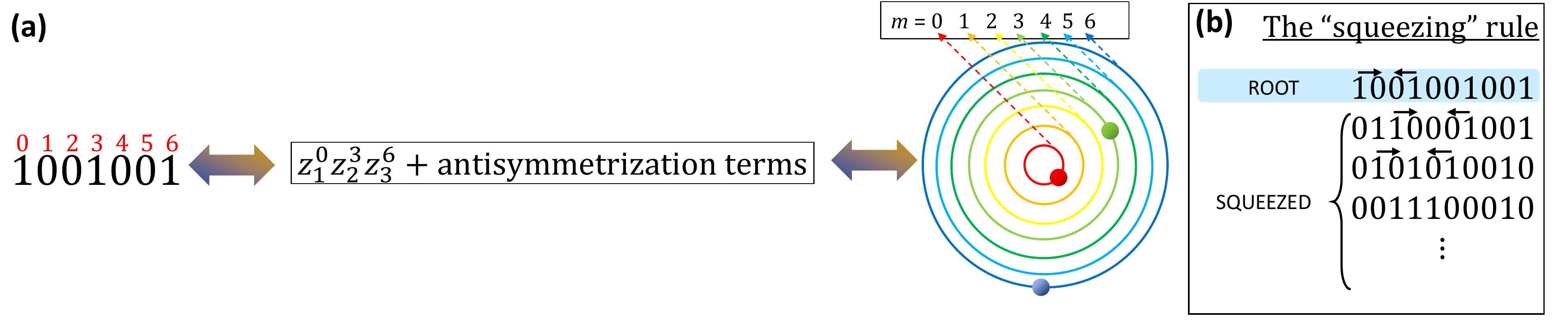}
\caption{(a) Correspondence between the binary representation of a configuration, the corresponding monomial in holonomic variable, and the electron occupation in orbitals on the disk. The single-particle wavefunction for the particle at orbital $m$ is $\psi_m\sim z^m$. (b) For any two partitions $\mu$ and $\nu$, we write $\mu<\nu$ if $\mu$ can be obtained from $\nu$ by moving two 1's in its representation closer by the same number of digits, as illustrated by the arrows above the configurations (the ``squeezing" rule). The Jack polynomial contains the monomials corresponding to the given root configuration and all the configuration squeezed from the root.}
\label{representation}
\end{center}
\end{figure*}

To illustrate what a ``root configuration" is, we start by a simple example of the Laughlin state with two particle. This is simply $(z_1-z_2)^3 = z_1^3-3z_1^2z_2+3z_1z_2^2-z_2^3=m_{\lambda}-3m_{\mu}$, where we let $m_{\lambda}=z_1^3-z_2^3$ and $m_{\mu}=z_1^2z_2-z_1z_2^2$. Note that both $m_{\lambda}$ and $m_{\mu}$ are anti-symmetric polynomials in $z_1$ and $z_2$. Before defining $\lambda$ and $\mu$, it is also instructive to look at the three-particle Laughlin state:
\begin{equation}
\label{Laughlin 3 particles}
\psi_{L}(z_1,z_2,z_3) = (z_1-z_2)^3(z_2-z_3)^3(z_1-z_3)^3
\end{equation}
Similar to the two-particle case, we can expand this product into a lot of terms. Rather than doing so in a brute-force manner, we note the followings: 

\begin{enumerate}[label=(\roman*)]
\item All terms in the sum are of the order 9. This means each term is of the form $z_1^{\lambda_1}z_2^{\lambda_2}z_3^{\lambda_3}$ where $\lambda_1+\lambda_2+\lambda_3=9$.
\item Since the wavefunction is anti-symmetric, for every term $z_1^{\lambda_1}z_2^{\lambda_2}z_3^{\lambda_3}$, there is also a term $\text{sgn}(\sigma)z_1^{\lambda_{\sigma(1)}}z_2^{\lambda_{\sigma(2)}}z_3^{\lambda_{\sigma(3)}}$ for $\sigma\in S_3$ and $\text{sgn}(\sigma)$ is the sign of the permutation $\sigma$. This means it is enough to capture all these anti-symmetric terms with a triplet $(\lambda_1,\lambda_2,\lambda_3)$ such that $\lambda_1<\lambda_2<\lambda_3$, which we call a partition. In general, a partition $\lambda$ of order $N$ is an $n$-tuplet $(\lambda_1,..,\lambda_n)$ such that $\lambda_i<\lambda_j$ for $i<j$ and $\lambda_1+...+\lambda_n=N$.

For a partition $\lambda$, we denote $m_\lambda$ the $n$-variable \emph{monomial} defined by
\begin{equation}
\label{monomial}
m_\lambda = \sum_{\sigma\in S_n}\text{sgn}(\sigma)z_1^{\lambda_{\sigma_1}}z_2^{\lambda_{\sigma_2}}...z_n^{\lambda_{\sigma_n}}
\end{equation}
We can represent a partition by a string of binary digits 0 and 1, where the indices of 1's, counted from the leftmost digit (and the leftmost digit has index $0$), corresponds to the partition terms. Fig. \ref{representation}a shows the correspondence between this binary representation of a partition and its monomial. We note also that on a disk, $z^m$ corresponds to the electron orbit of angular momentum $m$. Hence we can further identify the monomial with the arrangement of electrons in space. This way, going back to the partition, we can treat each digit in the binary representation as an orbital, where 0's represent empty orbitals and 1's represent orbitals filled by an electron (see Fig. \ref{representation}a).
\item Not all possible partitions of order 9 appear in $\psi_L$ in Eq. (\ref{Laughlin 3 particles}). In general, we find that if $m_\lambda$ appears in the wavefunction for some partition $\lambda$, then so does $m_\mu$ for a partition $\mu$ obtained from $\lambda$ by a ``squeezing" operation. This squeezing operation is best illustrated when writing the partition in binary format: it is done by taking two 1's and move them toward each other by the same number of steps (see Fig. \ref{representation}b). The squeezing operation defines a partial ordering - we write $\mu<\nu$ if $\mu$ can be obtained from $\nu$ by a series of squeezes. For each FQH state, there exists a \emph{maximally unsqueezed} partition, which we call the ``root configuration". This root configuration alone defines the FQH states, and there are physical information that can be immediately read off from it. For example, the root configuration for the Laughlin state is 1001001001..., from which we can see the filling factor is $\nu=1/3$, since there are one electron every three orbitals (there is a digit 1 for every three digits).
\end{enumerate}

An application of the root configuration is reading off the position of the quasiholes and deriving their dynamics. In the main text, we shows the following root configurations that correspond to the bound state and separate state of the Gaffnian quasihole pair:
\begin{align}
\d{{\ocirc{{\d{\blankchar}}}}}&\textsubring{0}\textsubring{1}00\textsubring{1}00\textsubring{1}00\textsubring{1}00\textsubring{1}00\textsubring{1}00\textsubring{1}00\textsubring{1}\textsubring{\color{white}1}\label{rootS1}\\
\d{\blankchar}{\ocirc{{\textsubring{\blankchar}}}}&\textsubring{0}1{\obullet{{0}}}1\textsubring{0}{\ocirc{{0}}}\textsubring{0}1{\obullet{{0}}}1\textsubring{0}{\ocirc{{0}}}\textsubring{0}1{\obullet{{0}}}1\textsubring{0}{\ocirc{{0}}}\textsubring{0}1{\obullet{{0}}}1\textsubring{0}\ocirc{{{\textsubring{\blankchar}}}}\d{\blankchar}\label{rootS2}
\end{align}
(Eq. (3) and (4) in main text).The location of the quasiholes can be read off from the root configuration. The Laughlin quasihole (LQ), marked by empty triangles above the configuration, is located where there is no electron in three consecutive orbitals. The Gaffnian quasiholes (GQ), on the other hand, occur where there are fewer than two electrons in five consectuve orbitals. They are marked with circles below the root configuration. The above description qualitatively show the positions of the Gaffnian quasihole pair that makes up the Laguhlin quasihole, marked in red crossed circles to distinguish from the background GQs that form the Laughlin ground state (empty circles).

To further illustrate quantitatively the positions of the quasiholes, we calculate the electron density at each orbital from each Jack polynomial. This is done by taking the corresponding Jack polynomials, $J_{01001001001..}^{\alpha=-2}$, for the bounded GQ pair at the north pole, and $J_{01010001010001..}^{\alpha=-3/2}$, for the separated GQs at the two poles. $J_{01001001001..}^{\alpha=-2}$ is also the Laughlin state. Each monomial in the Jack polynomial is squeezed from the root configuration - we can treat each term as a different electron distribution whose probablity is given by the square of its coefficient. The average electron density can then be calculated accordingly. 

\begin{figure}
\begin{center}
\includegraphics[width=\linewidth]{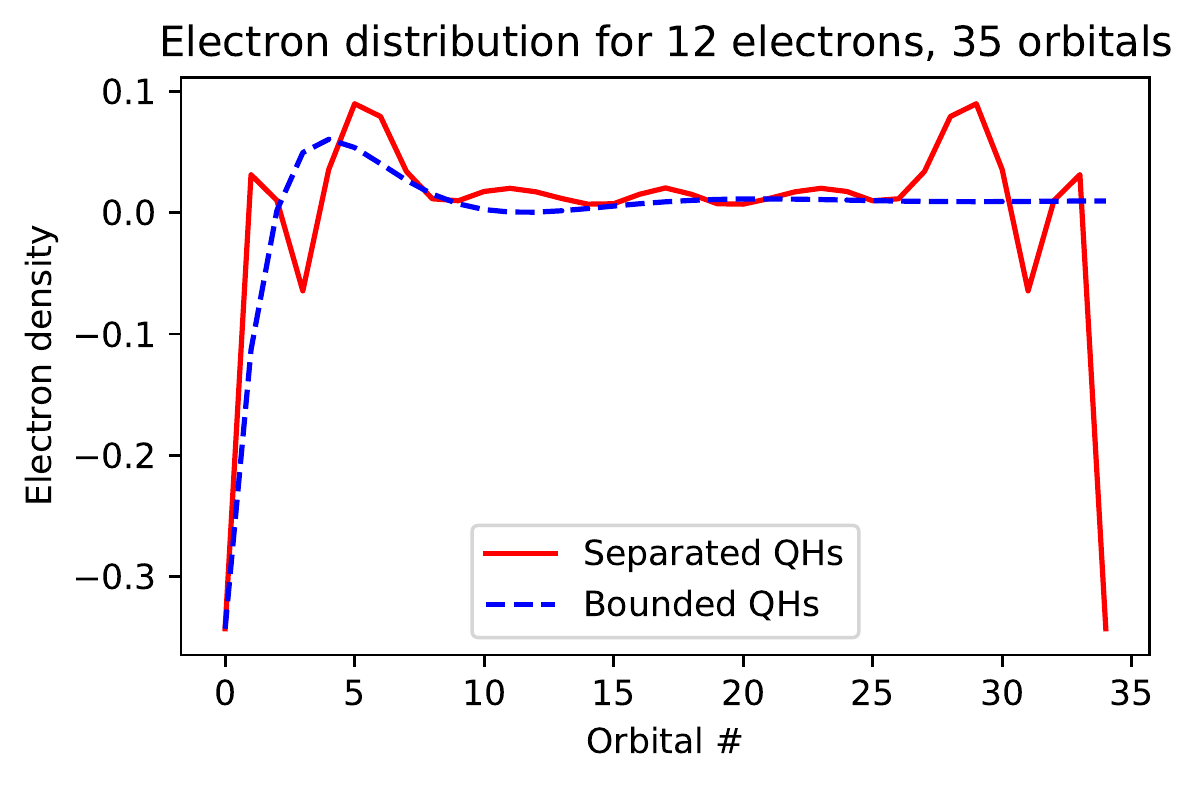}
\caption{Electron density calculated from $J_{01001001001..}^{\alpha=-2}$ (blue dashed line) and $J_{01010001010001..}^{\alpha=-3/2}$ (red solid line) for system with 12 electrons. The $y$-value is off-set by the average electron density (12/35) so that a negative value means a lack of electrons, i.e. a hole. Orbital number 0 corresponds to the north pole and orbital number 35 corresponds to the south pole on the sphere.}
\label{wf}
\end{center}
\end{figure}

Fig. \ref{wf} shows the average density per orbital, calculated for a system of 12 electrons and 35 orbitals. The $y$-axis is offset by the total average electron density (12/35) so that a negative value implies a hole. We see that the for the state with the root given in Eq.(\ref{rootS1}), the quasihole is strongly localized at the left-most orbital (north pole). For the state with the root given by Eq.(\ref{rootS2}), the wavefunction is more spread out, with two prevalent holes at either poles. We can also infer from the symmetric profile that the GQ pairs must be split.

\section{S2. Local exclusion constrain (LEC) formalism}
\begin{figure}
\begin{center}
\includegraphics[width=\linewidth]{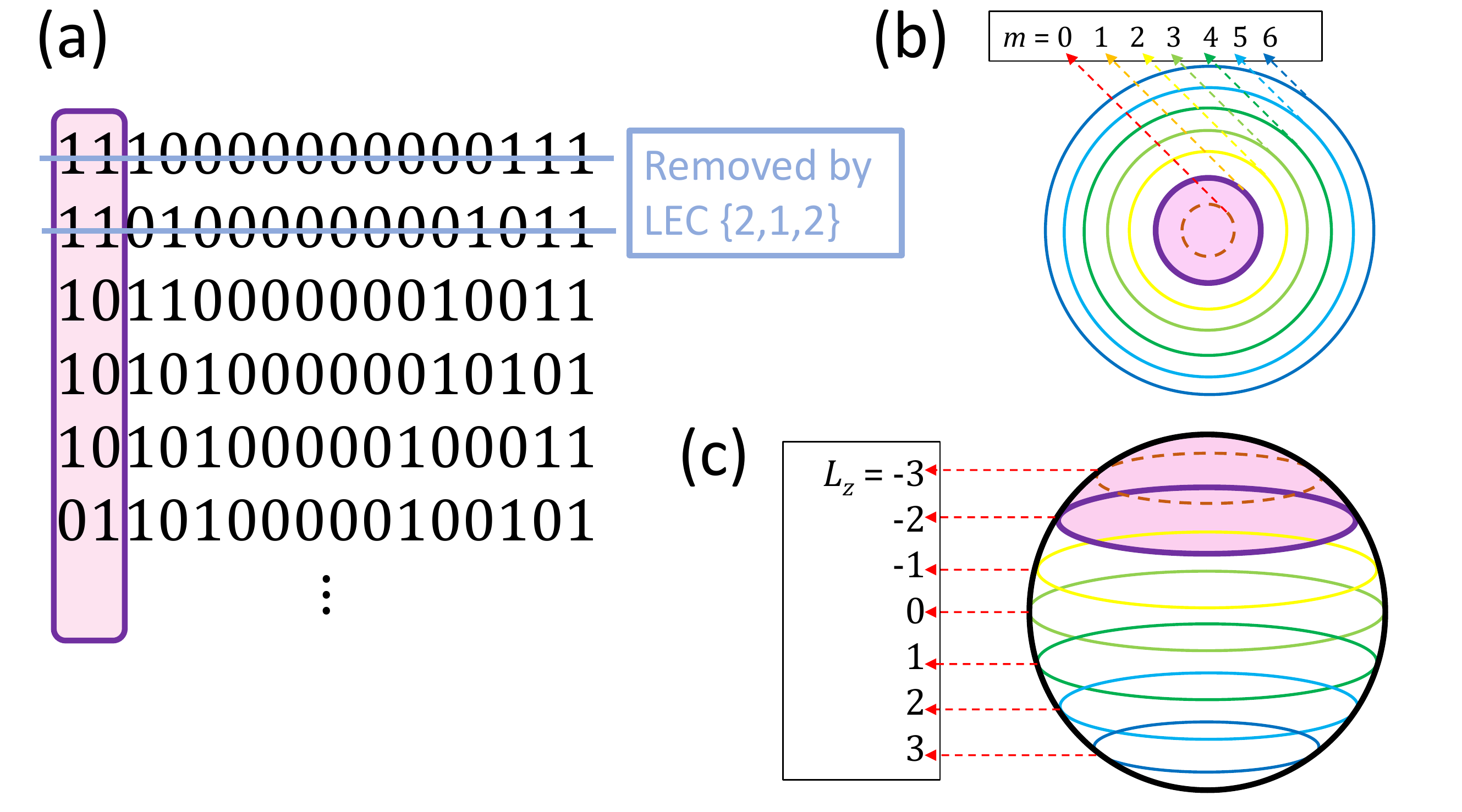}
\caption{(a) Truncation on the $L_z=0$ of the full Hilbert space with LEC \{2,1,2\}: we assert that within the left-most $n=2$ digits there are no more than $n_e=1$ digit 1's (b) The first two digits of a configuration corresponds to the droplet on the disk. Due to translational symmetry, this LEC is equivalent to a \emph{classical} constrain that we can observe no more than one electron within \emph{any} droplet of area $2\times2\pi\ell_B^2$. (c) On a spherical geometry, the left-most digits in the binary representation corresponds to a droplet at the north pole.}
\label{LEC}
\end{center}
\end{figure}

An LEC a triplet of integers $\hat{c}=\{n,n_e,n_h\}$ dictating what can or cannot be measured from a quantum state in a single Landau level. It is therefore a set of constraints on the reduced density matrix of the many-body FQH states. It asserts that within a given droplet of $n$ orbitals anywhere in the quantum fluid, there are no more than $n_e$ electrons and no more than $n_h$ holes (empty orbitals). In the monomial basis where each basis is represented by a binary string in the manner described above, we can impose LEC at the north pole by dictating that within the first $n$ orbitals, there are no more than $n_e$ digit 1's and no more than $n_h$ digit 0's. This serves as a truncation criteria for the Hilbert space. An example of truncation with LEC $\{2,1,2\}$ is illustrated in Fig. \ref{LEC}. Note that due to translational symmetry (rotational symmetry on a sphere), any constrain on a droplet at the origin (i.e. at the north pole on the Haldane sphere) is a constraint everywhere in the quantum fluid.

Given $N_o$ orbitals and $N_e$ electrons, we can construct the Hilbert space basis that consists of all binary sequences of $N_e$ 1's and $(N_o-N_e)$ 0's and remove all basis states that do not satisfy a given LEC at the north pole, which we denote as $\hat{c}$. Let the subspace spanned by the remaining states be $\mathcal{H}^t_{N_e,N_o}$. Within this subspace, we can find the \emph{highest weight} states which satisfy $\hat{L}^\dagger\ket{\psi}=0$; let $\bar{\mathcal{H}}^{\hat{c}}_{N_e,N_o}$ be the subspace spanned by those states. It is observed that for a given $N_e$, for some LEC $\hat{c}$ there exist a unique minimal orbital number $N_o^d$ such that dim$\bar{\mathcal{H}}^{\hat{c}}_{N_e,N_o^d}=1$ and dim$\bar{\mathcal{H}}^{\hat{c}}_{N_e,N_o}=0$ for $N_o<N_o^d$. For some choice of $\hat{c}$, this unique state is the model FQH state at some filling factor $p/q$, while $N_o^d$ also satisfies the corresponding commensurability condition $N_o^d=\frac{q}{p}(N_o+S_e)-S_\phi$, with some non-negative integers $S_e, S_\phi$ (these are the topological shifts of the FQH phase). For example, $\hat c=\{2,1,2\}$ corresponds to the Laughlin state at $N_o=3N_e-2$. Other FQH states discussed in the main text are the Gaffnian, where $\hat c=\{2,1,2\}\lor\{5,2,5\}$ and the Haffnian, given by $\hat c=\{2,1,2\}\lor\{6,2,6\}$. Here $\hat{c}_1\lor\hat{c}_2$ means that the state must satisfy \emph{either} $\hat{c}_1$ \emph{or} $\hat{c}_2$.

It is conjectured that most, if not all FQH states can be determined by the above procedure\cite{Yang2019}. On top of providing a framework for writing down model states independent of a Hamiltonian, this formalism also reveals connections between different FQH states. This is because each LEC not only determines the topological indices and the ground state of a particular FQH phase, it also defines the quasihole space of the same phase\cite{Yang2019a,yang2019effective}. In the main text and in the section below, we note that the Laughlin state can be described as a fluid of Gaffnian quasihole. This comes from an observation that $\bar{\mathcal{H}}^{\{2,1,2\}}_{N_e,N_o}$ is a subspace of $\bar{\mathcal{H}}^{\{2,1,2\}\lor\{5,2,5\}}_{N_e,N_o}$. The latter is also the Gaffnian quasihole subspace, so using the notation from the main text, we have $\mathcal H_G=\bar{\mathcal{H}}^{\{2,1,2\}\lor\{5,2,5\}}_{N_e,N_o}$. We have also shown that the low lying excitations of the Laughlin phase can be spanned by $\mathcal H_G$ up to very good approximations, even when the realistic interaction is far away from $\hat V_1^{\text{2bdy}}$. Physically, this means that while for interactions far from $\hat{V}_1^{\text{2bdy}}$, such as $\hat{V}_{1LL}$, the Laughlin state is dressed with neutral excitations, the low-lying states are still captured by (undressed) Gaffnian quasiholes (see Fig. \ref{fig1}).

\section{S3. Laughlin state described by GQs}
In this section, we discuss in detail how the Laughlin state can be viewed in terms of the Gaffnian quasiholes. The first thing to note about this picture is that the Gaffnian state ($\nu$=2/5) is denser than the Laughlin state ($\nu=1/3$). To be precise, we look at the commensurability conditions of both the states. Given the same number of electrons $N_e$, the number of orbitals in the Laughlin ground state, $N_o^L$, and in the Gaffnian ground state, $N_o^G$ can be expressed as
\begin{align}
    N_o^L &= 3N_e-2\label{laughlin}\\
    N_o^G &= \frac{5}{2}N_e-3\label{gaffnian}
\end{align}
This means that compared to the Gaffnian ground state, the Laughlin ground state has additional orbitals. Recall that adding one orbital to the Gaffnian ground state yields two Gaffnian quasiholes (GQs). We can therefore view the Laughlin ground state as a fluid of $N_{GQ}$ Gaffnian quasiholes where
\begin{equation}
    \label{GQ number}
    N_{GQ} = 2(N_o^L-N_o^G) = N_e+2
\end{equation}
The root configuration of the Jack polynomial corresponding to the Laughlin state is given by:
\begin{equation}
    \label{Laughlin root}
    \textsubring{\blankchar}\textsubring{1}00\textsubring{1}00\textsubring{1}00\textsubring{1}00\textsubring{1}00\textsubring{1}00\textsubring{1}00\textsubring{1}\textsubring{\blankchar}
\end{equation}
in the $L=0$ sector, implying it has rotational symmetry on the sphere (i.e. translational symmetry in the thermodynamic limit). The empty circles below the configuration denote the positions of the GQs, which is where there are fewer than two electrons in five consecutive orbitals. We see that in the bulk, the GQs are evenly spaced, with one GQ every three orbitals.

\begin{figure*}
\begin{center}
\includegraphics[width=0.75\linewidth]{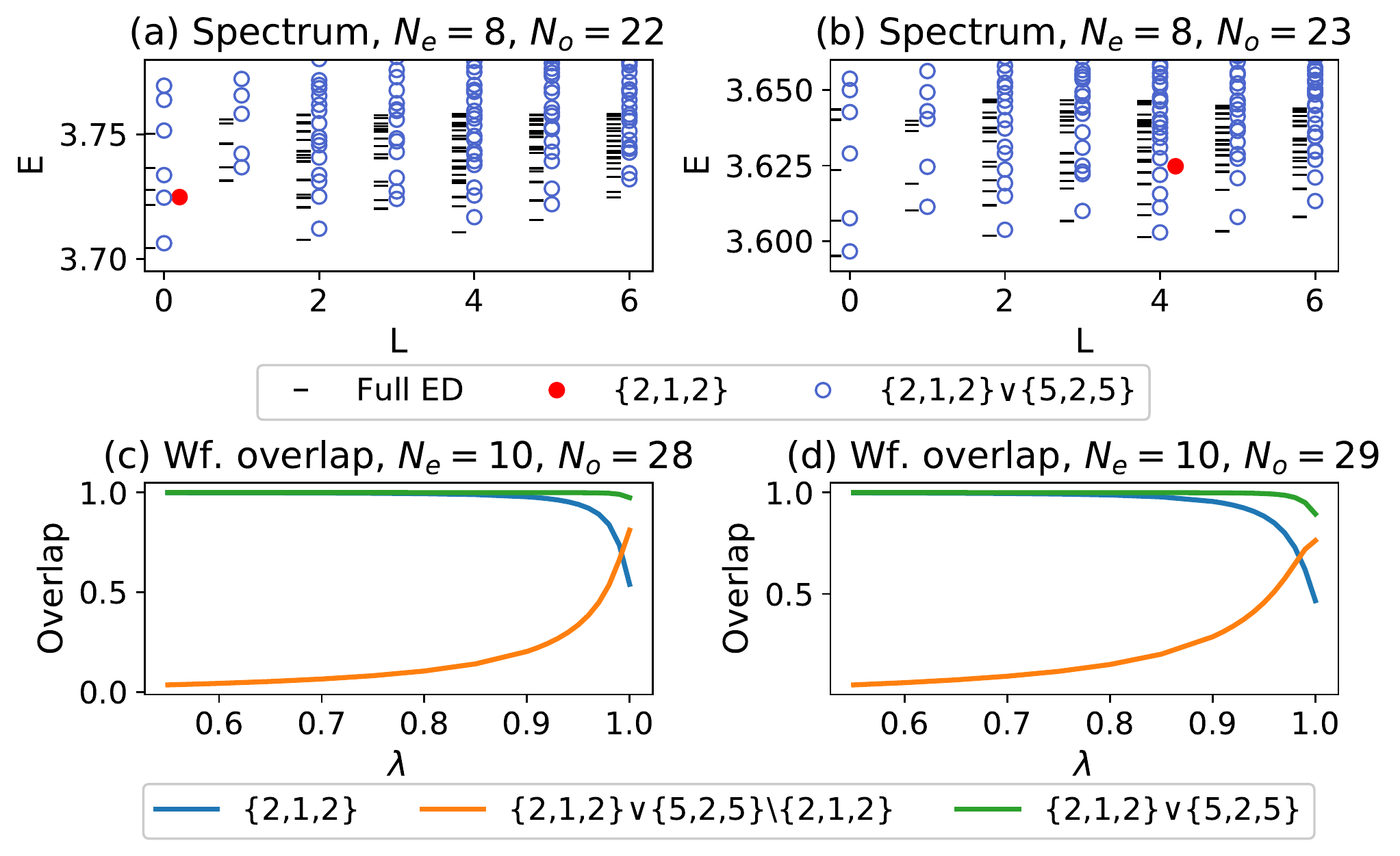}
\caption{\textit{Top row} -- Second Landau level spectrum with exact diagonalization (ED) on the full Hilbert space (black lines) and on the LEC subspaces $\{2,1,2\}$ (Laughlin state, red filled circles) and $\{2,1,2\}\lor\{5,3,5\}$ (GQs, blue empty circles) for: (a) ground state (b) one quasihole. \textit{Bottom row} -- Wavefunction overlap of the ground state of the ground state of $\hat H(\lambda)=(1-\lambda)\hat V_1^{\text{2bdy}}+\lambda \hat V_{\text{1LL}}$ with the Laughlin wavefunction (blue line), with the GQ subspace (green line), and the GQ subspace orthogonal to the Laughlin state (orange line) for: (c) ground state (d) one-quasihole state.}
\label{fig1}
\end{center}
\end{figure*}

This picture also explains an apparent paradox: we know that a GQ has a charge of $e/5$ with respect to the Gaffnian ground state. However, separating an LQ, which has a charge of $e/3$, yields two charges of $e/6$ each. The discrepancy is due to the fact that going from using electrons as the degree of freedom to the GQs, we have to account for the different background charge of the ground state. In the GQ picture, the Laughlin ground state has a charge density of $(e/5)\times(1/3)=e/15$, where $1/3$ is the density of the GQ. This is exactly the discrepancy between the additional single-GQ charge in the electron degree of freedom and that in the GQ degree of freedom. Thus with respect to the Laughlin ground state, each GQ carries a charge of $e/6$.

\begin{figure}
    \centering
    \includegraphics[width=\linewidth]{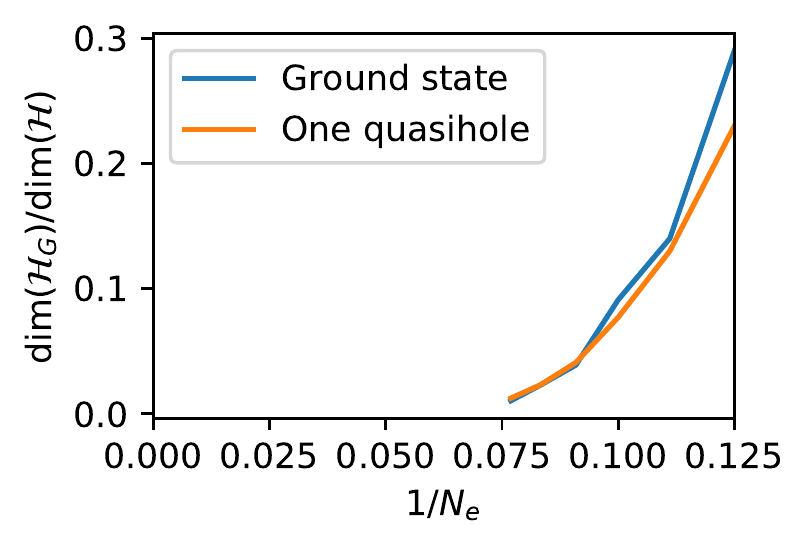}
    \caption{Finite size scaling of the ratio between the dimension of the Gaffnian sub-Hilbert space and the dimension of the full Hilbert space for the Laughlin ground state (blue line) and the one-quasihole state (orange line). }
    \label{ratio}
\end{figure}

The spectrum of the 1LL Coulomb interaction, $\hat V_{\text{1LL}}$, in $\mathcal H_G$ also supports the GQ description. Fig.\ref{fig1}a-b shows the spectrum obtained from numerics done on spherical geometry\cite{Haldane1983}. The variational energies from the subspace agree with the exact energies from the full Hilbert space very well for the low-lying states, in contrast to the energies from Laughlin model wavefunctions. Furthermore, we also note in the one-quasihole spectrum while the global ground state of $\hat V_1^{2bdy}$ resides in the $L_z=N_e/2$ sectors, the global ground state of $\hat V_{1LL}$ generally deviates from this $L_z$ sector ($L_z=0$ in Fig. \ref{fig1}b). This is also a result of $\hat V_3^{2bdy}$ favouring the split GQs.

\section{S4. Robustness and features of the spectrum near QCP}
In Fig. 2 of the main text, the interaction Hamiltonian was diagonalized in the Haffnian subspace $\mathcal{H}_H$ and its orthogonal complement in the Gaffnian subspace, $\mathcal{H}_G\setminus\mathcal{H}_H$. In doing this, we are able to separate the quadrupole excitation from the dipole excitation, and thus we can study the effect different interaction terms have on the individual branch. In particular, $\hat V_1^{2bdy}$ energetically favours the dipole branch, while $\hat V_H$ and $\hat V_3^{2bdy}$ favour the quadrupole branch.

\begin{figure*}
\begin{center}
\includegraphics[width=0.82\linewidth]{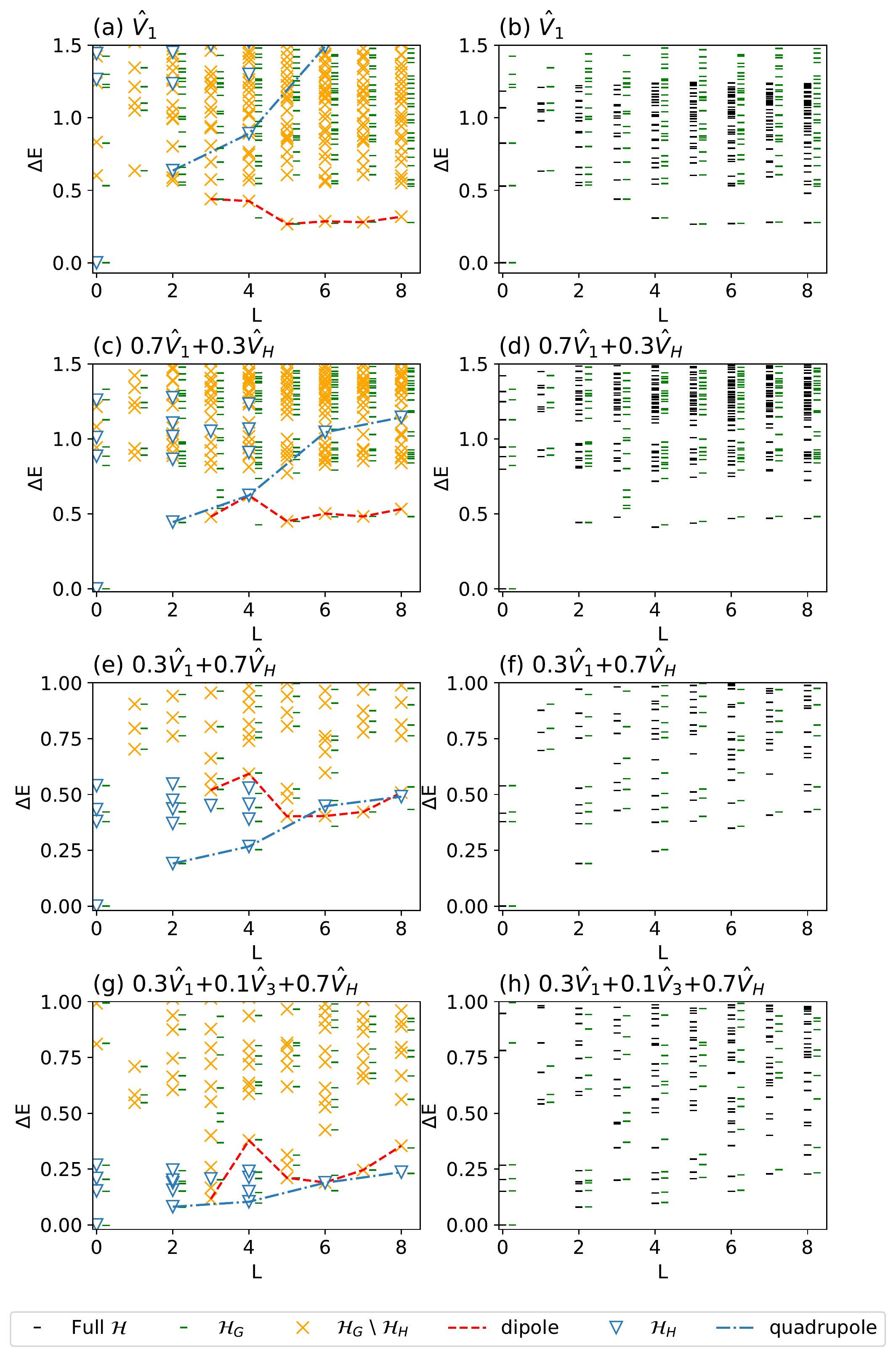}
\caption{Energy spectrum for system with 8 electrons and 22 orbitals obtained from exact diagonalization. \textit{Left column} -- The spectrum calculated individually in $\mathcal H_H$ and $\mathcal H_G\setminus\mathcal H_H$ match the spectrum calculated in $\mathcal H_G$. This indicates that $\mathcal H_H$ and its orthonormal complement $\mathcal H_G\setminus\mathcal H_H$ are also orthogonal with respect to the interactions. \textit{Right column} -- The spectrum calculated from the Gaffnian subspace (green lines, off-set to the right along the $x$-axis) also captures the low-lying energies of the spectrum in the full Hilbert space (black lines).  }
\label{full ed}
\end{center}
\end{figure*}

Here, we show that this result is robust with respect do different system size. Fig. \ref{full ed} shows the same spectrum for the system with 8 electrons and 22 orbitals. It can be easily seen that the same conclusion can be drawn. We also note the finite size effect particularly in the deviation of the quadrupole branch from a linear dispersion, expected of a Goldstone mode. This can also be understood in terms of the dynamics of the GQs, as the interaction amongs GQs are amplified in a smaller system.

Fig. \ref{full ed} also shows the diagonalization in the full Gaffnian subspace (green line, off-set in the $x$-axis). Although the Hamiltonian matrix is \textit{a priori} not block diagonal in $\mathcal{H}_H$ and $\mathcal{H}_G\setminus\mathcal{H}_H$, we see that the spectrum calculated in each of these subspace does not deviate significantly from the spectrum calculated in $\mathcal H_G$. This further illustrate our assertion that they describe two physically distinct pictures, namely one in which the GQs are separated ($\mathcal H_H$) and another in which the GQs are bounded in pairs ($\mathcal H_G\setminus\mathcal H_H$). The low-lying states in the spectrum in $\mathcal H_G$ also match the eneries of the low-lying states of the full spectrum well, as can be seen from the right column of Fig. \ref{full ed}. In particular, in the full spectrum we see the familiar magnetoroton mode (Fig. \ref{full ed}(b), starting from $L=3$ sector) slowly replaced by a linear excitation (Fig. \ref{full ed}(h), at $L=2,4,6,8$ sectors). These are the features that we successfully capture in the Gaffnian subspace (Fig. \ref{full ed}(a) and (g), respectively).

\section{S5. Separation energy of the Laughlin quasihole}
In this section, we give a formula for the separation energy of the Laughlin quasihole in terms of experimentally relevant quantities. To separate the GQ pair, i.e. going from Eq.(\ref{rootS1}) to Eq.(\ref{rootS2}), one can go through a series of root configurations as follow:
\begin{eqnarray}
&&\d{\blankchar}{\ocirc{{{\textsubring{\color{white}1}}}}}\textsubring{0}1{\obullet{{0}}}1\textsubring{0}{\ocirc{{0}}}\textsubring{0}\d{1}00\textsubring{1}00\textsubring{1}00\textsubring{1}00\textsubring{1}00\textsubring{1}\textsubring{\blankchar}\label{rootS3}\\
&&\d{\blankchar}{\ocirc{{{\textsubring{\blankchar}}}}}\textsubring{0}1{\obullet{{0}}}1\textsubring{0}{\ocirc{{0}}}\textsubring{0}1{\obullet{{0}}}1\textsubring{0}{\ocirc{{0}}}\textsubring{0}\d{1}00\textsubring{1}00\textsubring{1}00\textsubring{1}\textsubring{\blankchar}\label{rootS4}\\
&&\d{\blankchar}{\ocirc{{{\textsubring{\blankchar}}}}}\textsubring{0}1{\obullet{{0}}}1\textsubring{0}{\ocirc{{0}}}\textsubring{0}1{\obullet{{0}}}1\textsubring{0}{\ocirc{{0}}}\textsubring{0}1{\obullet{{0}}}1\textsubring{0}{\ocirc{{0}}}\textsubring{0}\d{1}00\textsubring{1}\textsubring{\blankchar}\label{rootS5}
\end{eqnarray}
In other words, separating the two GQs is achieved by creating Laughlin neutral excitations (NEs) between the them. It is from this observation that we say that the interaction between the two GQs are mediated by the Laughlin NEs. The dynamics of the GQs depend largely on the number of these NEs. 

In Fig. 2c-d in the main text, we see that the separation energy is (almost) proportional to the number of NEs between the GQs. Since the system is two-dimensional, the number of NEs is, in turn, proportional to the area, thus, we can write $\Delta E\propto \bar{d}^2$, or
\begin{equation}
\label{separation energy}
\Delta E = \bar\Delta_n \bar d^2
\end{equation}
where $\Delta E$ is the separation energy, $\bar d^2$ is the distance between the two GQs, and $\bar\Delta_n$ is a constant that we will deduce as following (after which the choice of notation will be clear).

Assuming that the GQs are evenly distributed in the system, the average distance between two GQs quasihole can be determined from the density of the GQs, $n_{\frac{e}{6}}$:
\begin{equation}
\label{distance}
\bar d = \frac{1}{\sqrt{n_{\frac{e}{6}}}}
\end{equation}

To determine $\bar\Delta_n$, we write
\begin{equation}
\label{delta n}
\bar\Delta_n = \frac{\Delta E}{\bar d^2}
\end{equation}
Consider on the sphere a state with a single separated GQ pair, each localized at a pole. The distance between them is equal to the length of the arc connecting two poles, given by
\begin{equation}
\label{arc length}
d = \pi R
\end{equation}
where $R$ is the radius of the sphere. $R$ can be expressed in terms of other variables\cite{greiter2011landau}:
\begin{equation}
\label{R}
R^2 = \frac{\hbar c s_0}{eB}
\end{equation}
where $s_0$ is the strength of the magnetic monopole at the center of the sphere, $e$ is the electron charge, and $B$ is the magnetic field strength on the surface of the sphere.

Next, we can also approximate $\Delta E$ in this case. We see from Eq. (\ref{rootS2}) that for a system with $N_e$ electrons, with one GQ at the north pole and the other at the south pole, the separation requires $N_e/2$ neutral excitations. Assuming that the NEs do not interact (as supported by numerical evidence), the energy cost for this is equal to $N_e/2$ times the energy required to create one neutral excitation. This can be approximated with $\Delta_n$, the neutral gap described in main text. This is a parameter that depends on the experimental details. In general, as seen from Fig. 3 of main text, the closer the system is to the nematic phase, the smaller $\Delta_n$ is.

Putting these information together, the separation energy on the sphere is approximately:
\begin{equation}
\label{sphere energy}
\Delta E \approx \frac{N_e}{2}\Delta_n = \frac{3N_o}{2}\Delta_n = \frac{3(2s_0+1)}{6}\Delta_n
\end{equation}
Here $N_o$ is the number of orbital, which equals to $3N_e$ in the Laughlin phase in the thermodynamic limit, and also to $2s_0+1$ on the sphere. Substituting the results in Eq.(\ref{arc length}), (\ref{R}), and (\ref{sphere energy}) to Eq.(\ref{delta n}) and taking the limit $s_0\rightarrow\infty$, we obtain
\begin{equation}
\label{delta n exp}
\bar\Delta_n = \frac{\Delta_n}{3\pi^2 l_B^2}
\end{equation}
We see that $\bar\Delta_n$ represents a ``reduced" neutral gap (hence the choice of notation). It depends on the neutral gap of the system and the strength of the external magnetic field, both of which are tuneable parameters in experiment.

Note that even though we rely on the spherical geometry for some of our calculation, the general dependence of $\Delta E$ on experimental parameter must be universal, with a possible scale factor in other geometries (e.g. disk). The separation energy depends on the number of GQs, the applied magnetic field, and the neutral gap of the system:
\begin{equation}
\label{separation energy exp}
\Delta E = \frac{\Delta_n}{3\pi^2 l_B^2}\frac{1}{n_{\frac{e}{6}}}
\end{equation}

\section{S6. Density of the Gaffnian quasiholes}
Knowing the separation energy allows us to approximate the densities of GQs and LQs in at finite temperature. We use a simplistic model, assuming each LQ is a two-level system: a bound state, and an unbound state with two separated GQs with energy $\Delta E$. Following the Boltzmann distribution, the ratio of density of GQs, $n_{\frac{e}{6}}$, to the density of LQs, $n$ is
\begin{equation}
\label{density ratio}
\frac{n_\frac{e}{6}}{n} = \frac{2e^{-\beta \Delta E}}{1+e^{-\beta \Delta E}}
\end{equation}
The factor of 2 is due to the fact that splitting one LQ yields two GQs. We define the reduced density,
\begin{equation}
\label{reduced density}
\bar n = \frac{n_{\frac{e}{6}}}{n}
\end{equation}
and the reduced temperature
\begin{equation}
\label{reduced temperature}
\bar\beta = \frac{\beta \bar\Delta_n}{n}
\end{equation}
We then get
\begin{equation}
\label{nbar equation}
\bar n = \frac{2e^{-\bar\beta/\bar{n}}}{1+e^{-\bar\beta/\bar{n}}}
\end{equation}
Solving Eq.(\ref{nbar equation}) gives the relative density of GQs to LQs in a system at finite temperature. This equation can also be simplified to
\begin{equation}
\bar n\left(1+e^{\bar\beta/\bar n}\right) = 2
\end{equation}
which is the same as Eq.(18) in the main text.

Let $f_{\bar\beta}(\bar n) =  \frac{2e^{-\bar\beta/\bar{n}}}{1+e^{-\bar\beta/\bar{n}}}$. The possible values of $\bar n$ are the solution to $\bar n = f_{\bar\beta}(\bar n)$. The solutions can be visualized as the intersection points of the graphs of $y=x$ and $y=f_{\bar\beta}(x)$, as shown in Fig. 4b in main text. We first note that $f_{\bar\beta}(0)=0$ for all $\bar\beta$, so there is always at least one solution. Either this is the only solution, in which case the system is in the regime of the Laughlin phase, where the Hamiltonian is closed to the model Hamitonian $V^{2bdy}_1$, or there can either be one or two other positive solutions (Fig. 4b, main text).

There exists a unique value of $\bar\beta$ at which there is only one non-zero value of $\bar{n}$ that satisfies (\ref{nbar equation}). Let this value be $\bar\beta_c$, and the non-zero solution be $\bar{n}_c$. Noting that under this condition, the graph of $y=f_{\bar\beta_c}(x)$ is tangential to $y=x$ at $x=\bar{n}$, we can determine $\bar\beta_c$ and $\bar{n}$ by solving the simultaneous equations:
\begin{align}
\bar n_c &= f_{\bar\beta_c}(\bar n_c)\label{eq1}\\
1&=f_{\bar\beta_c}'(\bar n_c)\label{eq2}
\end{align}
Solving this numerically yields $\bar\beta_c=0.557$ and $\bar n_c=0.436$.  A pair of non-zero solution exist for $\bar\beta<\bar\beta_c$. One solution approaches zero as $T\to\infty$. This is the ``trivial" case where the splitting of the GQ pair is not favourable and only LQs are observed in the system. The other solution approaches unity as $T\to\infty$. This is also favourable because the denser the GQs are in the system, the smaller the average distance between any two of them are, which lowers the variational energy. When a system prepared in the Laughlin regime is tuned past the critical point $\beta_c$, it spontaneously becomes a mixture of the two solutions.

From Eq. (\ref{reduced density}) and (\ref{reduced temperature}), we can express the reduced temperature in terms of physical quantities, noting that $n_{e/6}=\delta B/\Phi_0$ where $\delta B$ is the additional magnetic field and $\Phi_0=h/e$ is the flux quantum.
\begin{equation}
\label{reduced temperature 2}
\bar\beta = \frac{2T_n}{3\pi T}\left(\frac{\delta B}{B_0}\right)^{-1}
\end{equation}
where $B_0$ is the magnetic field strength at the center of the $1/3$ plateau, and $T_n=\Delta_n/k_B$ is the quadrupole gap temperature. This expression shows how the tuneable quantities $T$, $T_n$, and $\delta B$ can affect $\bar\beta$ and as a result affect the GQ density (see Fig. 4c in main text).

Eq. (\ref{reduced temperature 2}) can be inversed to express the temperature in terms of $\bar\beta$ and other variables. In particular, we find that the critical temperature is:
\begin{equation}
\label{critical temperature}
T_c = \frac{2T_n}{3\pi\bar\beta_c}\left(\frac{\delta B}{B_0}\right)^{-1}
\end{equation}
which is the same as Eq. (12) in main text.

\section{S7. Model Wavefunctions containing multiple quadrupole excitations}
For a many-body state containing one quadrupole excitation at $\nu=1/3$, the root configuration is given by:
\begin{eqnarray}
1100001001001001\cdots
\end{eqnarray}
There is a unique highest weight state containing only basis squeezed from this root configuration, using the scheme described in Ref.\cite{yang2012model}. Let this many-body state be $|\psi_{\text{qp}}\rangle$, it is the highest weight state uniquely determined by the following constraint:
\begin{eqnarray}
\hat V_1^{\text{2bdy}}\hat c_0\hat c_1|\psi_{\text{qp}}\rangle=0
\end{eqnarray}
where $\hat V_1^{\text{2bdy}}$ is the model Hamiltonian for the Laughlin state (the Haldane pseudopotential), while $\hat c_k$ annihilates an electron in the $k^{\text{th}}$ orbital from the north pole. Physically, this implies by removing the quadrupole excitation at the north pole, the state relaxes back to the Laughlin ground state, as it should be.

Many-body states containing multiple quadrupole excitations, e.g. Eq.(10) in the main text, can be similarly determined unambiguously. Let $|\psi_{\text{k-qp}}\rangle$ be the highest weight state containing $k$ quadrupole excitations at the north pole, it can thus be uniquely determined with the following constraint:
\begin{eqnarray}
\hat V_1^{\text{2bdy}}\hat c_0\hat c_1\hat c_4\hat c_5\cdots \hat c_{2k-2}\hat c_{2k-1}|\psi_{\text{qp}}\rangle=0
\end{eqnarray}
which is the state in the $L=2k$ sector. Again, the state relaxes back to the Laughlin ground state far away from the north pole, where all the quadrupole excitations are piled.

\bibliographystyle{apsrev}
\bibliography{ref_sm}

\begin{thebibliography}{59}
\bibitem[{foo({\natexlab{a}})}]{footnote4}
\bibinfo{note}{{A GQ is located at the center of any group of five consecutive
  orbitals with fewer than two electrons. An LQ is located at the center of any
  group of three consecutive orbitals with no electron. A Laughlin
  quasiparticle is located at the center of any group of three consecutive
  orbitals containing more than one electron.}}
\bibitem[{foo({\natexlab{b}})}]{footnote5}
\bibinfo{note}{{In particular we can now prove analytically the fractionalised
  Laughlin quasihole state and the quadrupole excitations are exact zero energy
  states of the Haffnian model Hamiltonian, and the results will be published
  soon.}}
\end{thebibliography}
\end{document}